\documentclass{article}

\usepackage{arxiv}
\usepackage[bottom]{footmisc}
\usepackage[utf8]{inputenc} 
\usepackage[T1]{fontenc}    
\usepackage{hyperref}       
\usepackage{url}            
\usepackage{booktabs}       
\usepackage{amsfonts}       
\usepackage{nicefrac}       
\usepackage{microtype}      
\usepackage{lipsum}
\usepackage{graphicx}
\usepackage{framed,multirow}
\usepackage{verbatim}
\usepackage{amssymb}
\usepackage{latexsym}
\usepackage{lineno}
\usepackage{color}
\graphicspath{{images/}}
\usepackage{amsmath,bm}
\usepackage{amsthm,psfrag}
\usepackage{amssymb}
\usepackage{mathtools}

\theoremstyle{definition}

\definecolor{dgreen}{rgb}{0.0,0.545,0.0}

\usepackage{floatrow}

\usepackage{array}
\newcolumntype{P}[1]{>{\centering\arraybackslash}p{#1}}
\usepackage{tabularx} 
\usepackage{caption}
\usepackage[labelfont=bf]{caption}
\usepackage{pifont}

\usepackage{psfrag}
\usepackage{tabularx}
\usepackage{makecell}
\usepackage{algorithm}
\usepackage{algpseudocode}
\usepackage{amsmath}

\usepackage{url}
\usepackage{xcolor}
\definecolor{newcolor}{rgb}{.8,.349,.1}

\title{RoeNets: Predicting Discontinuity of Hyperbolic Systems from Continuous Data}

\author{
  Shiying Xiong \thanks{corresponding author, email: shiying.xiong@dartmouth.edu} \\
  Department of Computer Science\\
  Dartmouth College\\
  Hanover, NH 03755 \\
  \And
  Xingzhe He \\
  Department of Computer Science\\
  Dartmouth College\\
  Hanover, NH 03755 \\
  \And
  Yunjin Tong  \\
  Department of Computer Science\\
  Dartmouth College\\
  Hanover, NH 03755 \\
  \And
  Runze Liu  \\
  Department of Computer Science\\
  Dartmouth College\\
  Hanover, NH 03755 \\
  \And
  Bo Zhu \\
  Department of Computer Science\\
  Dartmouth College\\
  Hanover, NH 03755 \\
}

\begin{document}
\maketitle
\begin{abstract}
We introduce Roe Neural Networks (RoeNets) that can predict the discontinuity of the hyperbolic conservation laws (HCLs) based on short-term discontinuous and even continuous training data. Our methodology is inspired by Roe approximate Riemann solver (P. L. Roe, J. Comput. Phys., vol. 43, 1981, pp. 357--372), which is one of the most fundamental HCLs numerical solvers. In order to accurately solve the HCLs, Roe argues the need to construct a Roe matrix that fulfills "Property U", including diagonalizable with real eigenvalues, consistent with the exact Jacobian, and preserving conserved quantities. However, the construction of such matrix cannot be achieved by any general numerical method. Our model made a breakthrough improvement in solving the HCLs by applying Roe solver under a neural network perspective. To enhance the expressiveness of our model, we incorporate pseudoinverses into a novel context to enable a hidden dimension so that we are flexible with the number of parameters. The ability of our model to predict long-term discontinuity from a short window of continuous training data is in general considered impossible using traditional machine learning approaches.
We demonstrate that our model can generate highly accurate predictions of evolution of convection without dissipation and the discontinuity of hyperbolic systems from smooth training data.
\end{abstract}

\section{Introduction}\label{sec:intro}

Hyperbolic conservation law (HCL) is a long-standing subject in the fields of magnetohydrodynamics \cite{Arthur2006}, hydrodynamics \cite{Bressan2011}, aerodynamic \cite{MaoKang2020}, combustion physics \cite{Terao1989}, and nuclear physics \cite{Scheid1974}. It is often challenging to develop advanced numerical HCL solvers that is capable of simultaneously resolving small-scale flow structures \cite{Harten1997,Park2010} and capturing discontinuities \cite{Ames1970}\cite{Burger2008}.

Generally speaking, numerical methods discretize the HCL problems into grids, which can be seen approximately as Riemann problems on a local scale \cite{Griebel1999} \cite{Colella2006}\cite{Vilar2011} \cite{McCorquodale2011} \cite{Spekreijse1987}. The Riemann problem is a hyperbolic partial differential equation (PDE), with initial data comprised of two constant states, separated by a single discontinuity. Many pieces of research have been devoted to the Riemann problem, since solutions to Riemann problems may give exact solutions to complex nonlinear equations, such as the Euler equations \cite{Toro1997}. Therefore, the development of mathematical theory of HCLs relies extensively on the understanding originated in studies of Riemann problems. 


Integrating the Riemann problem into numerically solving HCLs can be traced back to the work of Godunov \cite{godunov1959difference}.  Godunov's method is only first-order accurate in both space and time, but gives solutions that preserve monotonicity of the data. It is widely used as a base scheme for developing higher-order methods \cite{van1979towards}. While Godunov's method solves the Riemann problem exactly, most modern methods use approximate Riemann solvers for lower computational cost. The approximate Riemann solver can also be designed so as to make the overall numerical solver more robust, by avoiding physically irrelevant solution states.

In 1981, Phil Roe introduced the Roe solver, a linearized Riemann solver, which improves the performance of Godunov’s method (less cost and less dissipation). The Roe solver involves finding an estimate for the intercell numerical flux at the interface between two computational cells, on some discretised space-time computational domain. The elegance of the method lies in the fact that the linearization also preserves the non-linear behaviors such as shocks and contact discontinuities \cite{Roe1981}.

The Roe solver consists of finding a Roe matrix $\tilde{\bm A}$  that is assumed constant between two cells, and the construction of the Roe matrix has to fulfill "Property U" : being diagonalizable with real eigenvalues, being consistent with the exact Jacobian, and preserving conserved quantities. Insofar as current numerical methods, the construction of such matrix cannot be attained.   

We propose an effective approach to solve HCLs by applying Roe solver under a neural network perspective.
Given the diagonalization of the Roe matrix $\tilde{\bm A} = \bm {L}^{-1}\bm {\Lambda} \bm L$, our model consists of two networks which learn $\bm {L}$ and $\bm {\Lambda}$ respectively. Using neural networks to directly approximate $\bm {L}$ and $\bm {\Lambda}$ is ineffective, since the number of learning parameters is limited by the number of components. To enhance the expressiveness of our model, we apply pseudoinverses in a novel context by replacing $\bm {L}^{-1}$ with $\bm L^{+} = (\bm L^{T} \bm L)^{-1}\bm L^T $ to enable a hidden dimension so that we are flexible with the number of parameters.

Compared with a standard numerical Roe solver, our model exhibits higher accuracy and efficiency, as well as a much stronger expressive power. 
We examine our model’s ability to solve multiple first-order linear hyperbolic PDEs varying from one to three components and furthermore, a Riemann problem with non-linear flux function, the Sod shock tube problem. The results show that our model can successfully predict these problems with high accuracy and strong robustness. Most importantly, we show our model is capable of predicting discontinuity with smooth training data for Burgers’ equation without dissipation. The ability of our model to predict long-term discontinuity from a short window of continuous training data is in general considered impossible if using traditional machine learning approaches.  

In summary, we propose RoeNet, which makes the following contributions: 
\begin{itemize}
    \item Provide an effective approach to solving HCLs by applying Roe solver under a neural network perspective.
    \item Apply pseudoinverses in a novel context to enable a hidden dimension that enhance expressiveness of the neural networks.
    \item Exhibit high accuracy and strong robustness in solving HCLs with different number of components.
    \item Outperform Roe solver in better capturing the discontinuities in Riemann problems with both linear and nonlinear flux functions.
    \item Succeed in predicting long-term discontinuity from a short window of smooth training data.
\end{itemize}

\section{Background and related work}



\textbf{Riemann solvers}

Computing the numerical flux across a discontinuity in the Riemann problem is the primary goal of Riemann solvers. Typically the right and left states for the Riemann problem are calculated using some form of nonlinear reconstruction, such as a Total Variation Diminishing (TVD) Scheme \cite{Harten1997} or a WENO method \cite{liu1994weighted}, and then used as the input for the Riemann solver.  Some Riemann solvers other than the Godunov scheme \cite{godunov1959difference} and the Roe solver \cite{Roe1981} introduced previously include the HLL family of solvers. HLL stands for Harten, Lax, and van Leer, who first proposed
a method of this kind \cite{Harten1997}. The central idea is to assume a wave configuration for the solution that consists of two waves separating three constant states. Some various derivatives
of HLL solver are HLLE (Harten, Lax, Leer, and Einfeldt) solver \cite{Harten1997,einfeldt1988godunov} and HLLC (Harten-Lax-van Leer-Contact) solver \cite{toro1994restoration}. More recently, Rotated-hybrid Riemann solvers were introduced by Hiroaki Nishikawa and Kitamura, in order to overcome the carbuncle problems of the Roe solver and the excessive diffusion of the HLLE solver at the same time \cite{nishikawa2008very}.

\textbf{Deep learning solvers}

Approximating discontinuous functions with deep learning network has theoretical foundation in various literature, e.g., Yarosky's \cite{Yarosky2016} work on the Hölder space, Petersen and Voigtlaender \cite{Petersen2017} on piece-wise smooth functions, Imaizumi and Fukumizu \cite{Imaizumi2019} on DNN outperforming linear estimators and Suzuki's study \cite{Suzuki2018} on deep learning's higher adaptivity to spatial inhomogeneity of the target function. With above-mentioned theoretical cornerstone, a Physics Informed Neural Network (PINN) is proposed by Raissi, et al \cite{Raissi2017} to provide data-driven solutions to nonlinear problems, employing the well-known capacity of Deep Neural Networks (DNN) as universal function approximators \cite{Hornik1989}. Among its notable features, PINN maintains symmetry, invariance and conservation principles deriving from physical laws that governs observed data \cite{Zhang2019}. Michoski et al's work \cite{Michoski2019} show that without any regularization, irregular solutions to PDE can be captured. Mao et al. used PINN to approximate solutions to high-speed flows by formulating the Euler equation and initial/boundary conditions into the loss function\cite{Mao2020}. However, in Mao's setting, PINN does not solve the forward problems as accurately as the traditional numerical methods.  By incorporating invariants and data \emph{a priori} known to loss functions, such DNNs are also less adaptive to different kinds of problems.

\section{Methods}\label{sec:math} 
\subsection{Hyperbolic conservation laws}

A one dimensional HCL is a first-order partial differential equation (PDE)
of the form
\begin{equation}
\frac{\partial \bm u}{\partial t} + \frac{\partial \bm F(\bm u)}{\partial x} = 0,
\label{eq:conserv_u}
\end{equation}
with a initial condition
\begin{equation}
    \bm u(t=t_0,x) = \bm u_0(x),
    \label{eq:initial_u}
\end{equation}
and a proper boundary condition.
Here $\bm u = (u^{(1)},u^{(2)},\cdots,u^{(N_c)})$ with $N_c$ components is called the conserved quantity, while $\bm F =  (F^{(1)},F^{(2)},\cdots,F^{(N_c)})$ is the flux. The variable $t\in [t_0,t_T]$ denotes time, while $x\in \Omega$ is the space variable. 

We remark that for the discontinuous solution, \eqref{eq:conserv_u} is interpreted as a weak solution satisfying
\begin{equation}
    \int\int_{[t_0,t_T]\times\Omega}\bm u\frac{\partial \phi}{\partial t}+ \bm F \frac{\partial \phi}{\partial x} \textrm{d}t \textrm{d}x = 0,
\end{equation}
where $\phi$ is an arbitrary test function with a smooth and compact support.

In addition, \eqref{eq:conserv_u} can be written in a high dimensional form
\begin{equation}
\frac{\partial \bm u}{\partial t} + \sum_{i=1}^{N_d}\frac{\partial \bm F_i(\bm u)}{\partial x_i}=\bm 0.
\label{eq:conserv_u_highdim}
\end{equation}
If we can successfully solve \eqref{eq:conserv_u}, \eqref{eq:conserv_u_highdim} can be solved spontaneously by applying the method of approximating $\partial \bm F(\bm u)/\partial x$ to approximate $\partial \bm F_i(\bm u)/\partial x_i$.

\subsection{Roe solver}
The Roe solver \cite{Roe1981} discretizes \eqref{eq:conserv_u} as
\begin{equation}
\bm{u}_j^{n+1}=\bm{u}_j^{n} - \lambda_r\left(\hat{\bm F}_{j+\frac12}^n - \hat{\bm F}_{j-\frac12}^n\right),
\label{eq:un1}
\end{equation}
where $\lambda_r = \Delta t/\Delta x$ is the ratio of the temporal step size $\Delta t$ to the spatial step size $\Delta x$; $j=1,...,N_g$ is the grid node index; and 
\begin{equation}
    \hat{\bm F}_{j+\frac12}^n = \hat{\bm F}(\bm u_j^n,\bm u_{j+1}^n)
    \label{eq:Fj}
\end{equation}
with
\begin{equation}
    \hat{\bm F}(\bm u, \bm v) = \frac{1}{2}\left[\bm F(\bm u)+ \bm F(\bm v)-|\tilde{\bm A}(\bm u, \bm v)|(\bm v - \bm u)\right].
    \label{eq:Fuv}
\end{equation}
Here, Roe matrix $\tilde{\bm A}$ that is assumed constant between two cells,
and must obey the following Roe conditions (termed property U):
\begin{enumerate}
    \item Diagonalizable with real eigenvalues: ensures that the new linear system is truly hyperbolic.
    \item Consistency with the exact Jacobian: when $\bm u_j,\bm u_{j+1} \rightarrow \bm u$, we demand that $\tilde{\bm A}(\bm u_j, \bm u_{j+1}) = \partial \bm F (\bm u)/ \partial x$.
    \item Conserving $\bm F_{j+1}-\bm F_{j}=\tilde{\bm  A}(\bm u_{j+1}-\bm u_{j})$.
\end{enumerate}

From the first Roe condition, matrix $\tilde{\bm A}$ can be diagonalized as 
\begin{equation}
    \tilde{\bm A} = \bm {L}^{-1}\bm {\Lambda} \bm L. 
\label{eq:diag}
\end{equation}
Therefore, $|\tilde{\bm A}(\bm u, \bm v)|$ can be interpreted as
\begin{equation}
    |\tilde{\bm A}| = \bm {L}^{-1}|\bm {\Lambda}| \bm L. \label{eq:absA}
\end{equation}
Substituting \eqref{eq:Fj}, \eqref{eq:Fuv} and \eqref{eq:absA} into \eqref{eq:un1} along with the third Roe condition yields
\begin{equation}
\begin{aligned}
    \bm{u}_j^{n+1}=&\bm{u}_j^{n} - \frac{1}{2}\lambda_r[\bm L^{-1}_{j+\frac{1}{2}}(\bm \Lambda_{j+\frac{1}{2}}-|\bm \Lambda_{j+\frac{1}{2}}|)\bm L_{j+\frac{1}{2}}(\bm u_{j+1}^n-\bm u_{j}^n)\\
    &+\bm L^{-1}_{j-\frac{1}{2}}(\bm \Lambda_{j-\frac{1}{2}}+|\bm \Lambda_{j-\frac{1}{2}}|)\bm L_{j-\frac{1}{2}}(\bm u_{j}^n-\bm u_{j-1}^n)],
\end{aligned}
 \label{eq:roeeq}
\end{equation}
with
\begin{equation}
    \bm L_{j+\frac{1}{2}}^n = \bm L(\bm u_{j}^n,\bm u_{j+1}^n),~~~~ \bm \Lambda_{j+\frac{1}{2}}^n = \bm \Lambda(\bm u_{j}^n,\bm u_{j+1}^n).
\end{equation}
\eqref{eq:roeeq} serves as a template of evolution from $\bm{u}_j^{n}$ to $\bm{u}_j^{n+1}$.

In order to construct a Roe matrix $\tilde{\bm A}$ that follows the Roe conditions, Roe solver utilizes an analytical approach to solve $\bm L$ and $\bm \Lambda$ based on $\bm F(\bm u)$. The Roe matrix is then plugged into \eqref{eq:roeeq} to ultimately solve for $\bm u$ in \eqref{eq:conserv_u}. 
The Roe solver made a `smart' linearization of the Riemann problem, which is computationally efficient while still recognizing the non-linear jumps in the problem. Compared with the other Riemann solvers, e.g. Godunov’s method \cite{godunov1959difference} and HLLC solver \cite{toro1994restoration}, it performs with less cost and less dissipation. 

\subsection{Neural network architecture}
There are several inherent constraints of the Roe solver. First, it can only construct $\bm L$ and $\bm \Lambda$ for very limited number of flux functions $\bm F(\bm u)$. Given a $\bm F(\bm u)$, constructing $\bm L$ and $\bm \Lambda$ that fulfill property U is difficult no matter through numerical or analytical approach. Second, even with a constructed Roe matrix, the solution of \eqref{eq:conserv_u} still cannot be found accurately, since Roe solver does not offer a solution to find the best Roe matrices amongst all the possible matrices that fulfill property U. To tackle this challenges, while preserving the inherent numerical advantages of the Roe solver, we aim to develop a machine learning method that is capable of solving \eqref{eq:conserv_u} given any arbitrary flux function with both high efficiency and accuracy.
A naive design choice of a neural network is to directly approximate $\bm L$ and $\bm \Lambda$, which is ineffective, however, due to the limited number of learning parameters by the number of components $N_c$ in \eqref{eq:roeeq}. 

To solve this problem, we incorporate pseudoinverses into a novel context to enable a hidden dimension $N_h$ that could be much larger (or smaller) than $N_c$. Specifically, we replace $\bm L^{-1}$ in \eqref{eq:roeeq} with $(\bm L^{T} \bm L)^{-1}\bm L^{T} $, which is the pseudoinverse of $\bm L$ \cite{Pal1994}. Pseudoinverse, or Moore–Penrose inverse \cite{Ben-Israel1980}\cite{Rao1972} is well studied to produce least-square optimal learning, to compute generic vector-Jacobian products used in automatic differentiation\cite{chen2018neural}. By having pseudoinverses, we create a hidden dimension so that we are flexible with the number of parameters. This enhances the expressive ability of our model by a great extent. Therefore, we define the 
neural-network version of (\ref{eq:roeeq}) by replacing the inverse of $\bm L$ with its pseudoinverse
\begin{equation}
\begin{aligned}
    \bm{u}_j^{n+1}=&\bm{u}_j^{n} - \lambda_r \bm L^{+}_{j+\frac{1}{2}}(\bm \Lambda_{j+\frac{1}{2}}-|\bm \Lambda_{j+\frac{1}{2}}|)\bm L_{j+\frac{1}{2}}(\bm u_{j+1}^n-\bm u_{j}^n)\\
    & - \lambda_r \bm L^{+}_{j-\frac{1}{2}}(\bm \Lambda_{j-\frac{1}{2}}+|\bm \Lambda_{j-\frac{1}{2}}|)\bm L_{j-\frac{1}{2}}(\bm u_{j}^n-\bm u_{j-1}^n),
\end{aligned}
 \label{eq:roenet}
\end{equation}
where $\bm L^{+} = (\bm L^{T} \bm L)^{-1}\bm L^{T}$ is the pseudoinverse of $\bm L$.

\begin{figure}
  \centering
  \includegraphics[width=1.\linewidth]{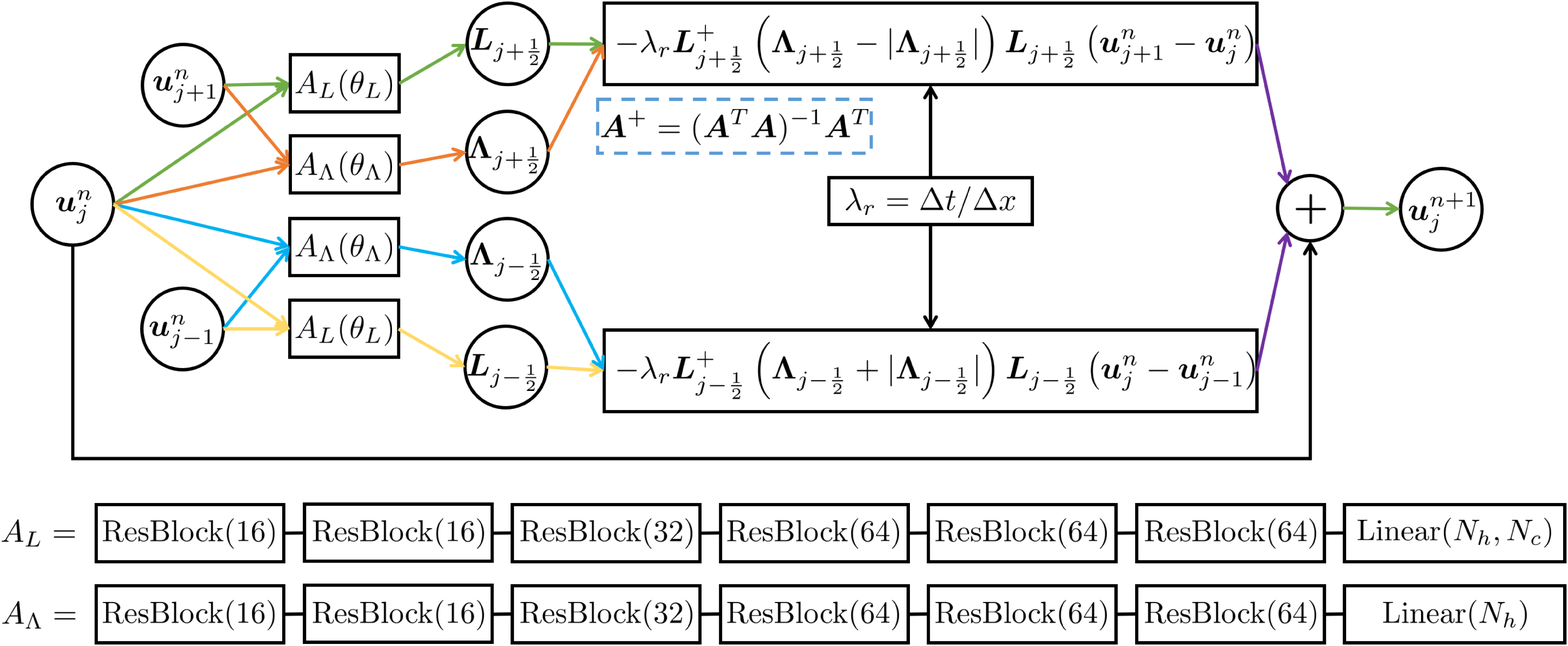}\\[0.5mm]
  \caption{The architecture of RoeNet. RoeNet takes the current conserved quantity $\bm u^n_j$ and its direct neighbors, $\bm u^n_{j-1}$ and $\bm u^n_{j+1}$, as the input, and outputs the next conserved quantity $\bm u^{n+1}_{j}$. RoeNet contains two parts, each consists of a $A_L$ and a $A_{\Lambda}$. The first part takes $\bm u_j^n$ and $\bm u_{j-1}^n$ as input of both $A_L$ and $A_\Lambda$ and outputs $\bm L_{j-\frac{1}{2}}$ through $A_L$ and $\bm \Lambda_{j-\frac{1}{2}}$ through $A_\Lambda$. The second part takes $\bm u_j^n$ and $\bm u_{j+1}^n$ as input of both $A_L$ and $A_\Lambda$ and outputs $\bm L_{j+\frac{1}{2}}$ through $A_L$ and $\bm \Lambda_{j+\frac{1}{2}}$ through $A_\Lambda$. The outputs are combined through \eqref{eq:roenet} to obtain $\bm u_j^{n+1}$. $\bm A^{+}$ represents the pseudoinverse of $\bm A$ as $\bm A^{+}= (\bm A^{T}\bm A)^{-1}\bm A^T$, shown in the blue dashed box. $\lambda_r$ is the ratio of the temporal step size to the spatial step size. The ResBlock has the same architecture as in \cite{He2015resnet} only with the 2D convolution layers replaced by linear layers. The numbers in the parentheses are output dimensions of each Resblock.}
  \label{fig:RoeNet}
\end{figure}

Overall, RoeNet is constructed with two networks $A_L$ and $A_{\Lambda}$ , which learn $\bm L$ and $\bm \Lambda$ in \eqref{eq:diag} respectively. As shown in Figure \ref{fig:RoeNet}, RoeNet takes $\bm u^n_j$ and its direct neighbors, $\bm u^n_{j-1}$ and $\bm u^n_{j+1}$, as the input, and outputs $\bm u^{n+1}_{j}$. 

Specifically, RoeNet contains two parts, each consists of a $A_L$ and a $A_{\Lambda}$. The first part takes $\bm u_{j-1}^n$ and $\bm u_{j}^n$ as input of both $A_L$ and $A_\Lambda$ and outputs $\bm L_{j-\frac{1}{2}}$ through $A_L$ and $\Lambda_{j-\frac{1}{2}}$ through $A_\Lambda$. The input $\bm u_{j-1}^n$ and $\bm u_{j}^n$ is a vector $(\bm u^{n, (1)}_{j-1}, \cdots, \bm u^{n, (N_c)}_{j-1}, \bm u^{n, (1)}_{j},\cdots, \bm u^{n, (N_c)}_{j})$ of length $2N_c$. The output matrix $L_{j-\frac{1}{2}}$ is of size $(N_c\times N_h)$, and the other output matrix $\bm \Lambda_{j-\frac{1}{2}}$ is a diagonal matrix of size $(N_h\times N_h)$. The second part takes $\bm u_j^n$ and $\bm u_{j+1}^n$ as input of both $A_L$ and $A_\Lambda$ and outputs $\bm L_{j+\frac{1}{2}}$ through $A_L$ and $\bm \Lambda_{j+\frac{1}{2}}$ through $A_\Lambda$. The input $u_j^n$ and $u_{j+1}^n$ is a vector $(\bm u^{n, (1)}_{j},\cdots, \bm u^{n, (N_c)}_{j}, \bm u^{n, (1)}_{j+1}, \cdots, \bm u^{n, (N_c)}_{j+1})$ of length $2N_c$. The output matrices $\bm L_{j+\frac{1}{2}}$ and  $\bm \Lambda_{j+\frac{1}{2}}$ take the same form as the output matrices in the first part. Given the four output matrices $\bm L_{j-\frac{1}{2}}$, $\bm \Lambda_{j-\frac{1}{2}}$, $\bm L_{j+\frac{1}{2}}$, and  $\bm \Lambda_{j+\frac{1}{2}}$, we combine them through \eqref{eq:roenet} to obtain $\bm u_j^{n+1}$.

$A_L$ and $A_{\Lambda}$ both consist of a chain of ResBlock \cite{He2015resnet} with a linear layer at the end of size $(N_h\times N_c)$ and $(N_h)$, respectively. The $N_h$ numbers learned by $A_{\Lambda}$ is transferred into a diagonal matrix of $(N_h\times N_h)$ with the learned numbers as its diagonal. The ResBlock has the same architecture as in \cite{He2015resnet} only with the 2D convolution layers replaced by linear layers. The numbers in the parentheses are output dimensions of each Resblock. 

Note that although we only show the calculation for grid cell $j$, the process is the same for grid cells. Since each node are calculated independent from the others except its closest neighbors, we train them in parallel to achieve high efficiency.

In addition, to address different boundary conditions, we implement two ways of padding. For periodic boundary conditions, we use the periodic padding, e.g., if $j=0$, $\bm u_{j-1} = \bm u_{N_g}$, where $N_g$ is the number of grid node. For Neumann boundary conditions, we use the replicate padding, e.g., if $j=0$, then we set $\bm u^n_{j-1}=\bm u^n_0$.

\section{Experiments}

\begin{table}
  \caption{Experimental set-up of four PDE problems.}
  \centering
  \setlength{\tabcolsep}{3mm}{
  \begin{tabular}{lcccc}
  \hline
  &1C Linear&3C Linear&Sod tube& Inviscid Burgers \\
  \hline
Boundary condition &periodic &Neumann&Neumann&periodic \\
Time step $\Delta t$&0.01&0.001&0.001&0.001\\
Space step $\Delta x$&0.01&0.005&0.005&0.01\\
Training time span&0.1&0.02&0.02&0.001\\
Predicting time span&2&0.2&0.1&0.3\\
Dataset samples &100&2000&2000&100\\
Dataset generation & analytical &analytical &analytical &
$2^{\textrm{nd}}$ central difference\\ 
Components number $N_c$&1&3&3&1\\
Hidden dimension $N_h$&1&16&32&64\\
  \hline
  \end{tabular}}
  \label{tab:problems}
\end{table}
We examine our model's ability of solving different first-order linear hyperbolic PDEs and a Riemann problem with a nonlinear flux function, the Sod shock tube problem. Most importantly, we show our model is capable of predicting discontinuity with smooth training data for inviscid Burgers’ equation. The details of the parameters we set and important quantities about hidden layers can be found in Table \ref{tab:problems}. Note that for the last problem, inviscid Burgers’ equation, we use $2^{\textrm{nd}}$ central difference to generate dataset, since there is no analytical solution for Burgers’ equation. For all the problems, the range of $x$ we aim to solve are from $-0.5$ to $0.5$.

For all experiments, we use the Adam optimizer \cite{Adam2014} with a learning rate 0.001. The learning rate decays with a ratio of 0.9 for every 5 epochs. We use a batch size of 16 for all experiments. We choose the Mean Squared Error as our loss function for all experiments. All the models are trained for 100 epochs and converge in less than 5 minutes in a single Nvidia RTX 2080Ti GPU.

\subsection{First-order linear hyperbolic PDEs}\label{sec:firstlinear}

We first show our model's ability of predicting the results of different first-order linear hyperbolic PDEs of the form \eqref{eq:conserv_u} with $\bm F(\bm u) = \bm A \bm u$ and with one or multiple components. Here $\bm A$ denotes a $N_c\times N_c$ constant matrix.

\begin{figure}
  \psfrag{x}[c][c]{\small $x$}
  \psfrag{u}[c][c]{\small $u$}
  \psfrag{m}{\small RoeNet}
  \psfrag{p}{\small RoeNet (noise)}
  \psfrag{n}{\small Roe solver}
  \psfrag{l}{\small Exact}
  \psfrag{a}{\small (a)}
  \psfrag{b}{\small (b)}
  \psfrag{c}{\small (c)}
  \psfrag{d}[c][c]{\small $\lambda_u$}
  \centering
  \includegraphics[width=.333\linewidth]{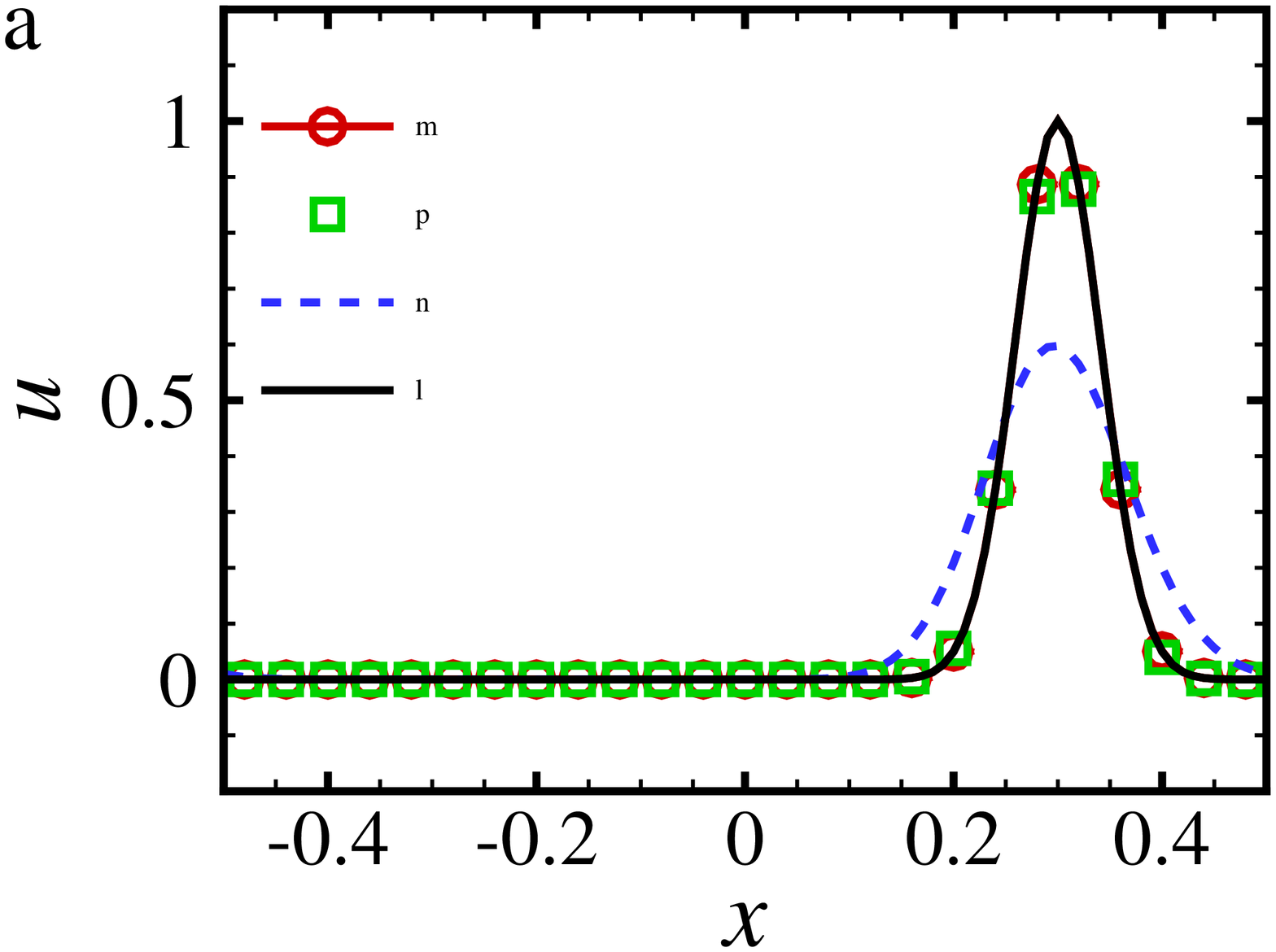}\hfill
  \includegraphics[width=.333\linewidth]{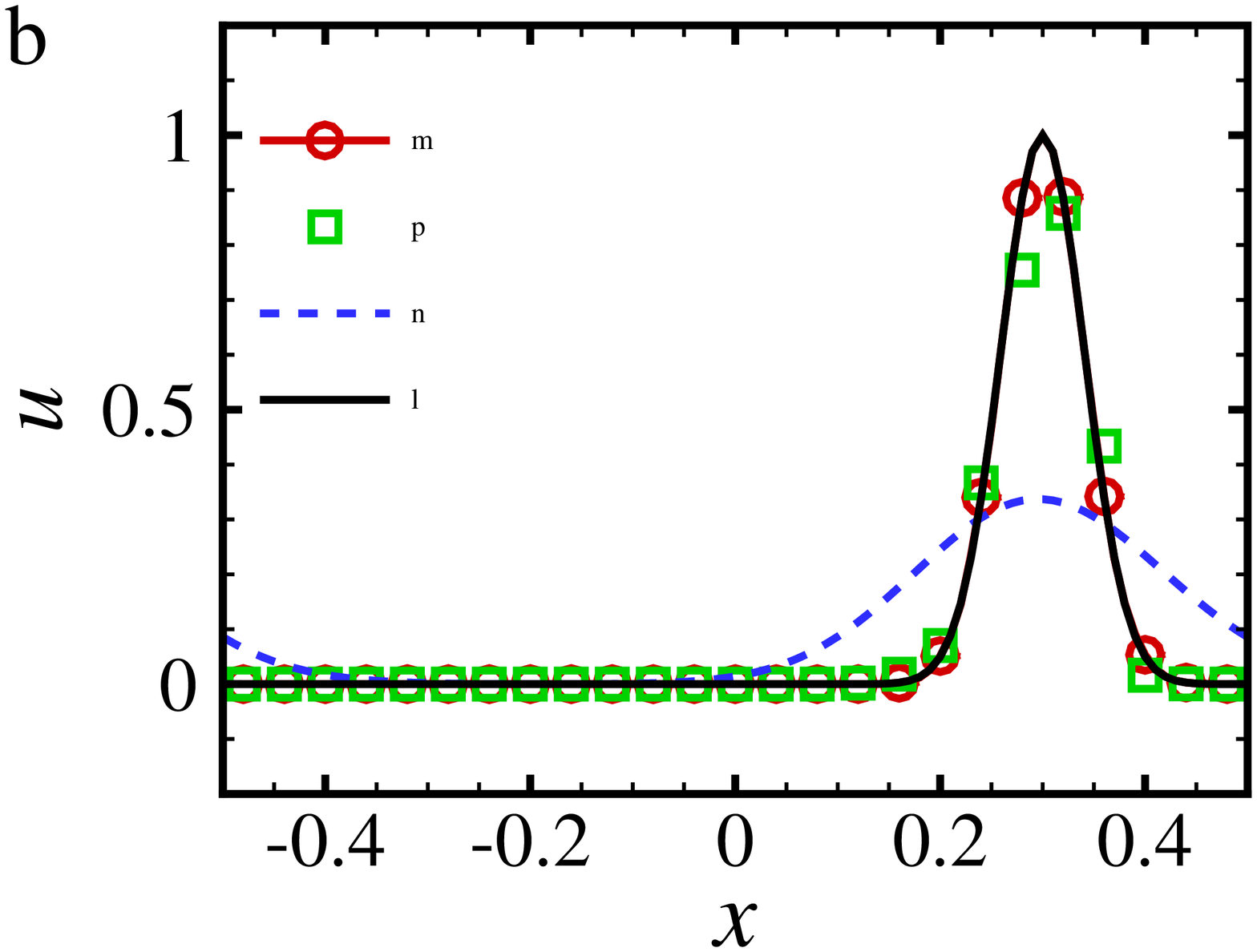}\hfill
  \includegraphics[width=.333\linewidth]{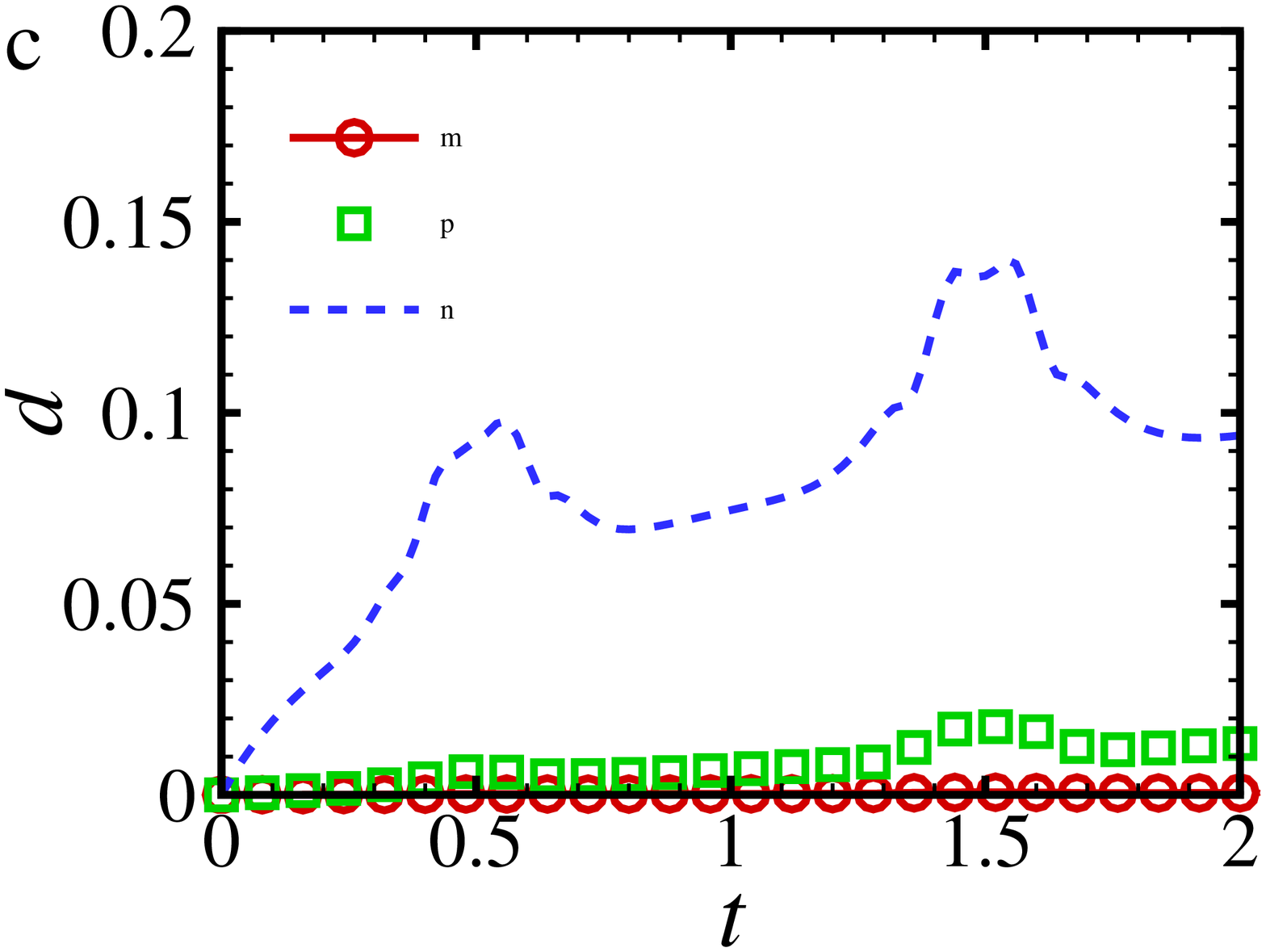}\\[0.5mm]
  \caption{Comparison of RoeNet and Roe solver for solving a one component linear hyperbolic PDE. (a) $t = 0.3$, (b) $t = 1.3$, and (c) averaged deviation $\lambda_u=\left<|u-u_{\textrm{exact}}|\right>$ of the predicted solutions from the exact solution. The legend "RoeNet (noise)" denotes the RoeNet with training noise $\epsilon \sim \mathcal{N}(0,0.1)$.}
  \label{fig:trivial1c}
\end{figure}

Figure \ref{fig:trivial1c} shows the predicting results of the linear hyperbolic PDE with one component (1C Linear)
\begin{equation}
\begin{dcases}
\bm  F =  x,\\
u(t=0,x) = e^{-300x^2}.
\end{dcases}
\label{eq:case1}
\end{equation}
In Figure \ref{fig:trivial1c} (a), we plot the results using RoeNet, RoeNet with noisy training data, and Roe solver, as well as the exact solution at $t=0.3$. It is clear that RoeNet outperforms the 
numerical Roe solver even when RoeNet is trained with noise $\epsilon \sim \mathcal{N}(0,0.1)$. At larger $t$, the predictions made by RoeNet with or without noise stay accurate, while the performance of the 
numerical Roe solver is getting worse, shown in Figure \ref{fig:trivial1c} (b). Figure \ref{fig:trivial1c} (c) shows the averaged deviation $\lambda_u=\left<|u-u_{\textrm{exact}}|\right>$ of the predicted solutions from the exact solution, where $\left<\cdot\right>$ denotes the average over $[-0.5,0.5]$. The averaged deviation of RoeNet indicated by the red circle line in Figure \ref{fig:trivial1c} (c) is almost negligible, showing the high accuracy of the prediction results made by RoeNet. The fact that the predicting error of RoeNet even with noise is more than 10 times smaller than that of numerical Roe Solver shows the high accuracy and strong robustness of RoeNet.

In addition, we apply RoeNet to solve a linear hyperbolic PDE with three components (3C Linear)
\begin{equation}
\begin{dcases}
    \bm F =  \begin{bmatrix}
    0.3237& 2.705 &5.4101\\
    0.3597& -0.4388&-2.8777\\
    -0.0144&0.0576&1.1151
    \end{bmatrix}\bm x,\\
    \bm u(t=0,x\leq 0) = (0.4,0.4,0.4),~~\bm u(t=0,x>0) = (-0.4,-0.4,-0.4).
\end{dcases}
\label{eq:case2}
\end{equation}

\begin{figure}
  \psfrag{x}[c][c]{\small $x$}
  \psfrag{u}[c][c]{\small $u^{(1)}$}
  \psfrag{v}[c][c]{\small $u^{(2)}$}
  \psfrag{w}[c][c]{\small $u^{(3)}$}
  \psfrag{m}{\small RoeNet}
  \psfrag{n}{\small Roe solver}
  \psfrag{l}{\small Exact}
  \psfrag{a}{\small (a)}
  \psfrag{b}{\small (b)}
  \psfrag{c}{\small (c)}
  \centering
  \includegraphics[width=.333\linewidth]{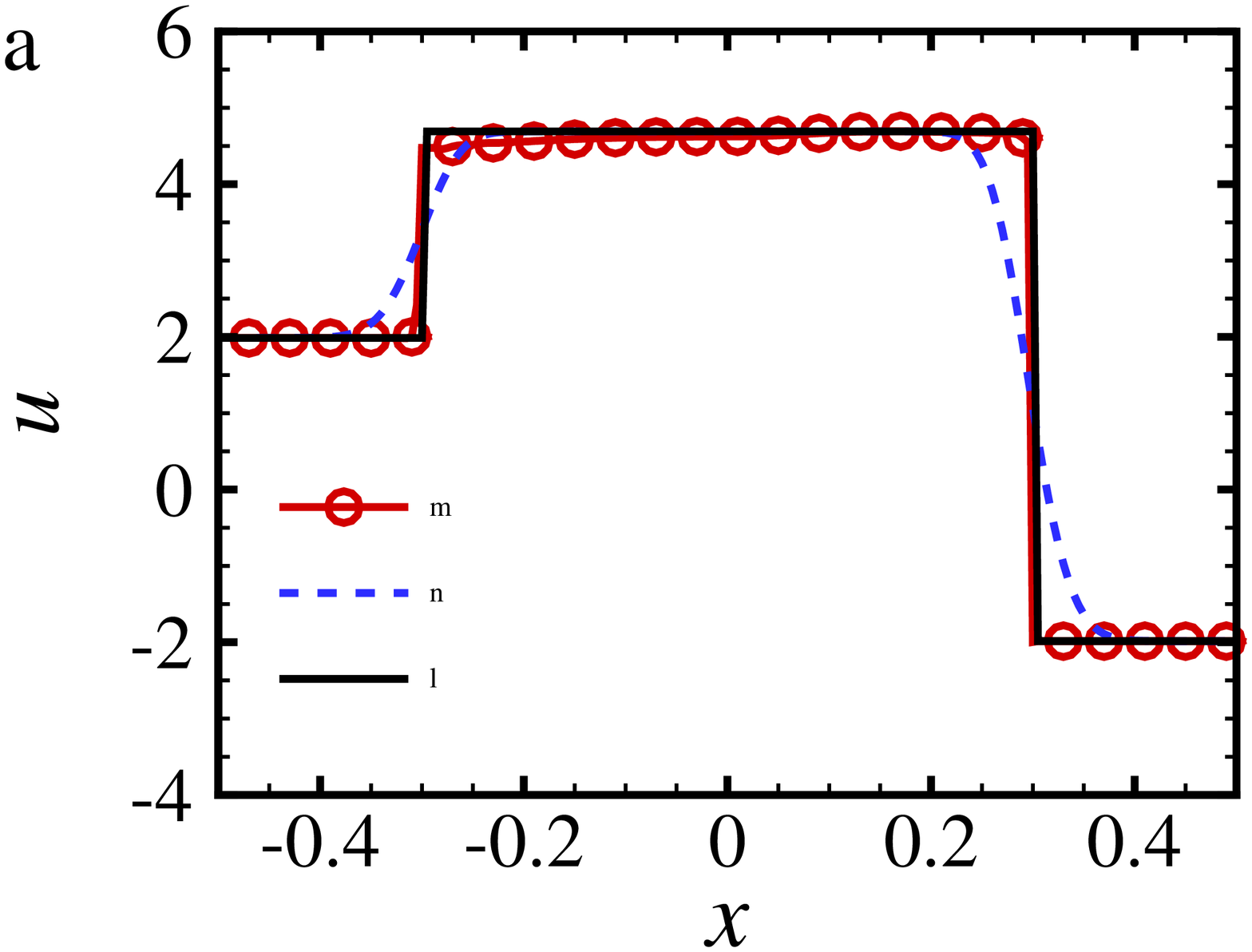}\hfill
  \includegraphics[width=.333\linewidth]{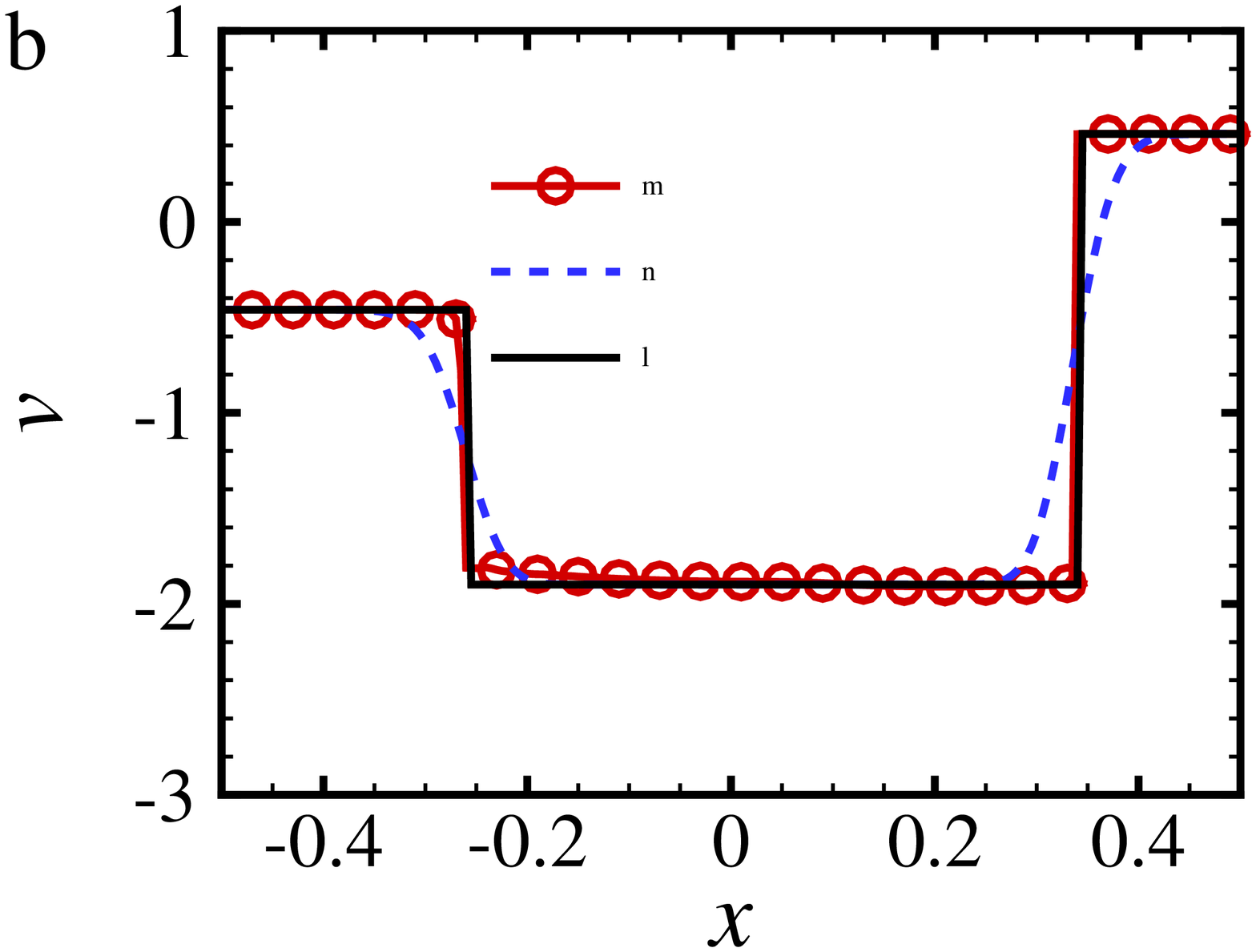}\hfill
  \includegraphics[width=.333\linewidth]{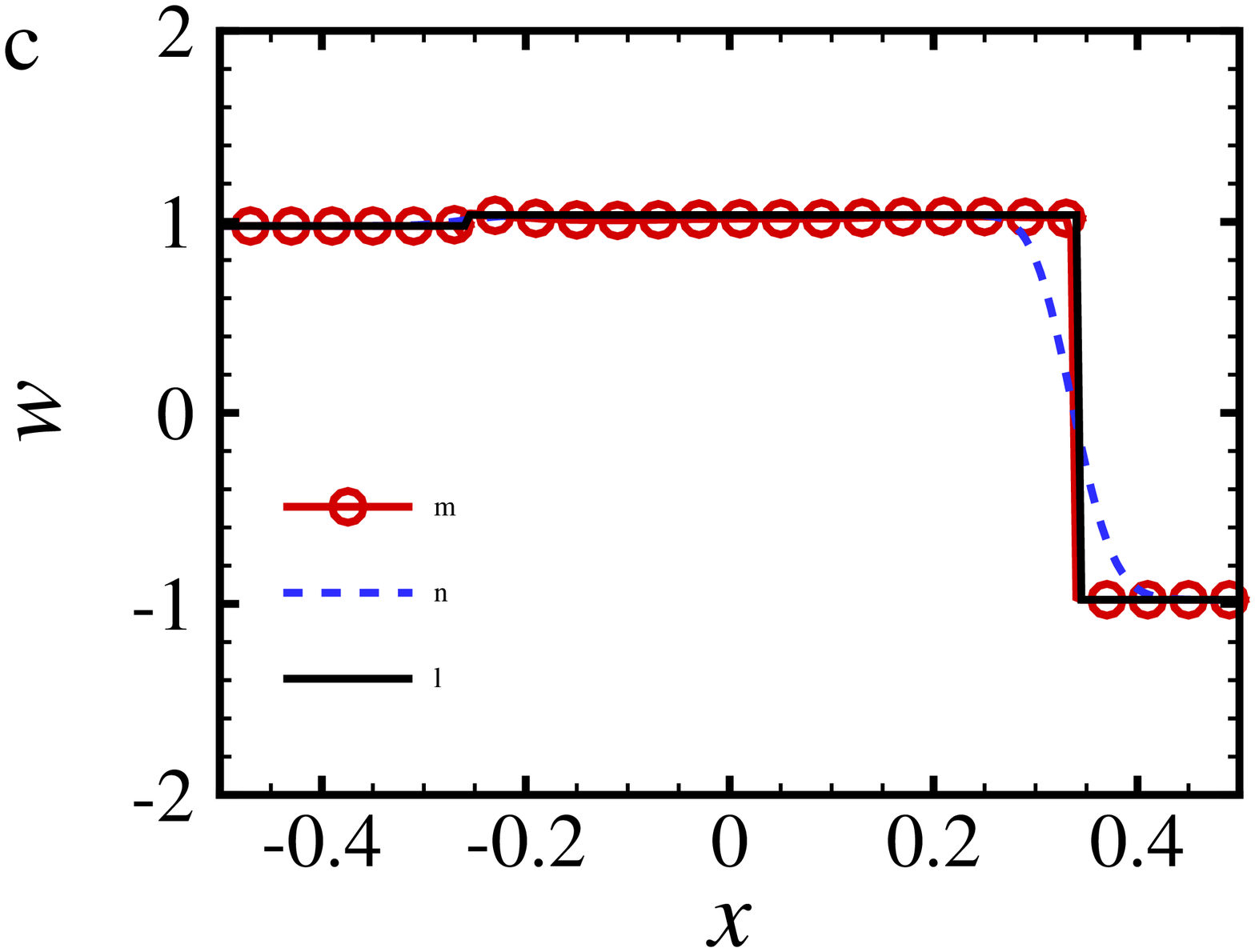}\\[0.5mm]
  \caption{Riemann problem with three components and linear flux function. (a), (b), and (c) plot the prediction results using RoeNet and Roe solver, and exact solutions of the three components $u^{(1)}$, $u^{(2)}$, and $u^{(3)}$ respectively.}
  \label{fig:trivial3c}
\end{figure}

Figure \ref{fig:trivial3c} shows the the exact solutions and the prediction results of the three components $u^{(1)}$, $u^{(2)}$, and $u^{(3)}$ of a Riemann problem with linear flux function. From all three plots in Figure \ref{fig:trivial3c}, we can observe that the predictions made by RoeNet match the exact solutions perfectly, while these of Roe solver have obvious errors around the discontinuous points (at $x \approx \pm0.3$).

\subsection{Riemann problems with nonlinear flux functions}
\begin{figure}
  \psfrag{x}[c][c]{\small $x$}
  \psfrag{u}[c][c]{\small $u^{(1)}$}
  \psfrag{v}[c][c]{\small $u^{(2)}$}
  \psfrag{w}[c][c]{\small $u^{(3)}$}
  \psfrag{m}{\small RoeNet}
  \psfrag{n}{\small Roe solver}
  \psfrag{l}{\small Exact}
  \psfrag{a}{\small (a)}
  \psfrag{b}{\small (b)}
  \psfrag{c}{\small (c)}
  \centering
  \includegraphics[width=.333\linewidth]{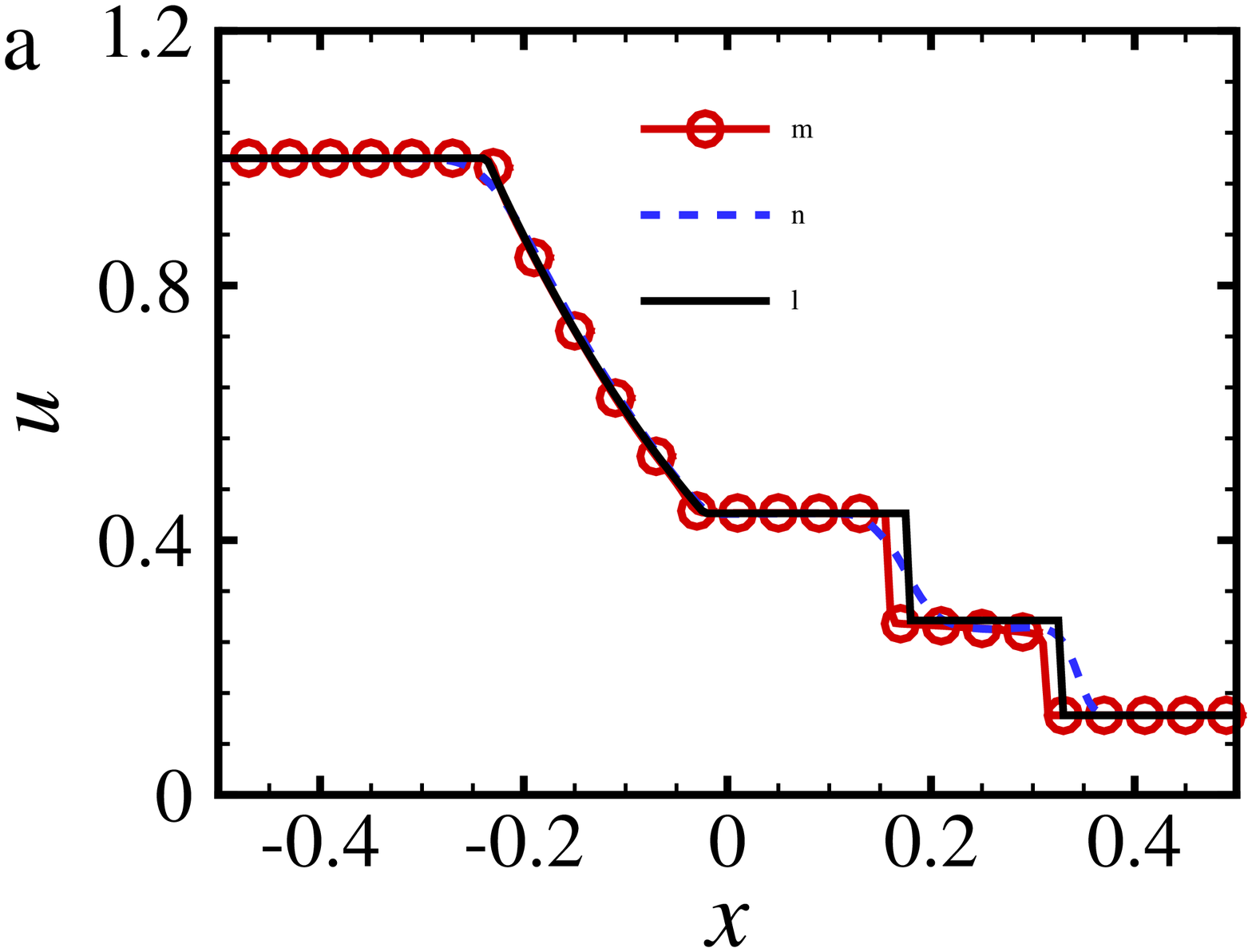}\hfill
  \includegraphics[width=.333\linewidth]{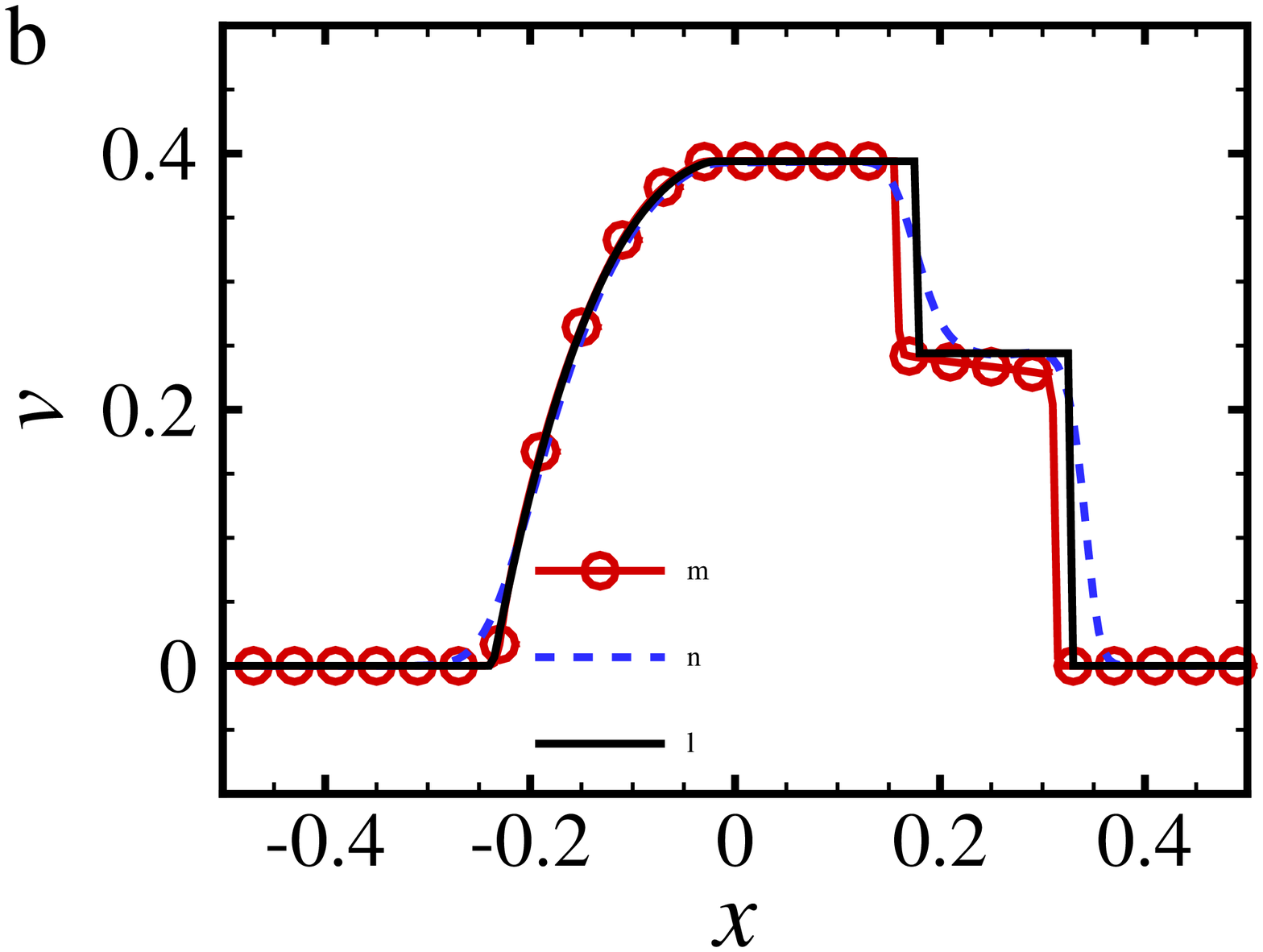}\hfill
  \includegraphics[width=.333\linewidth]{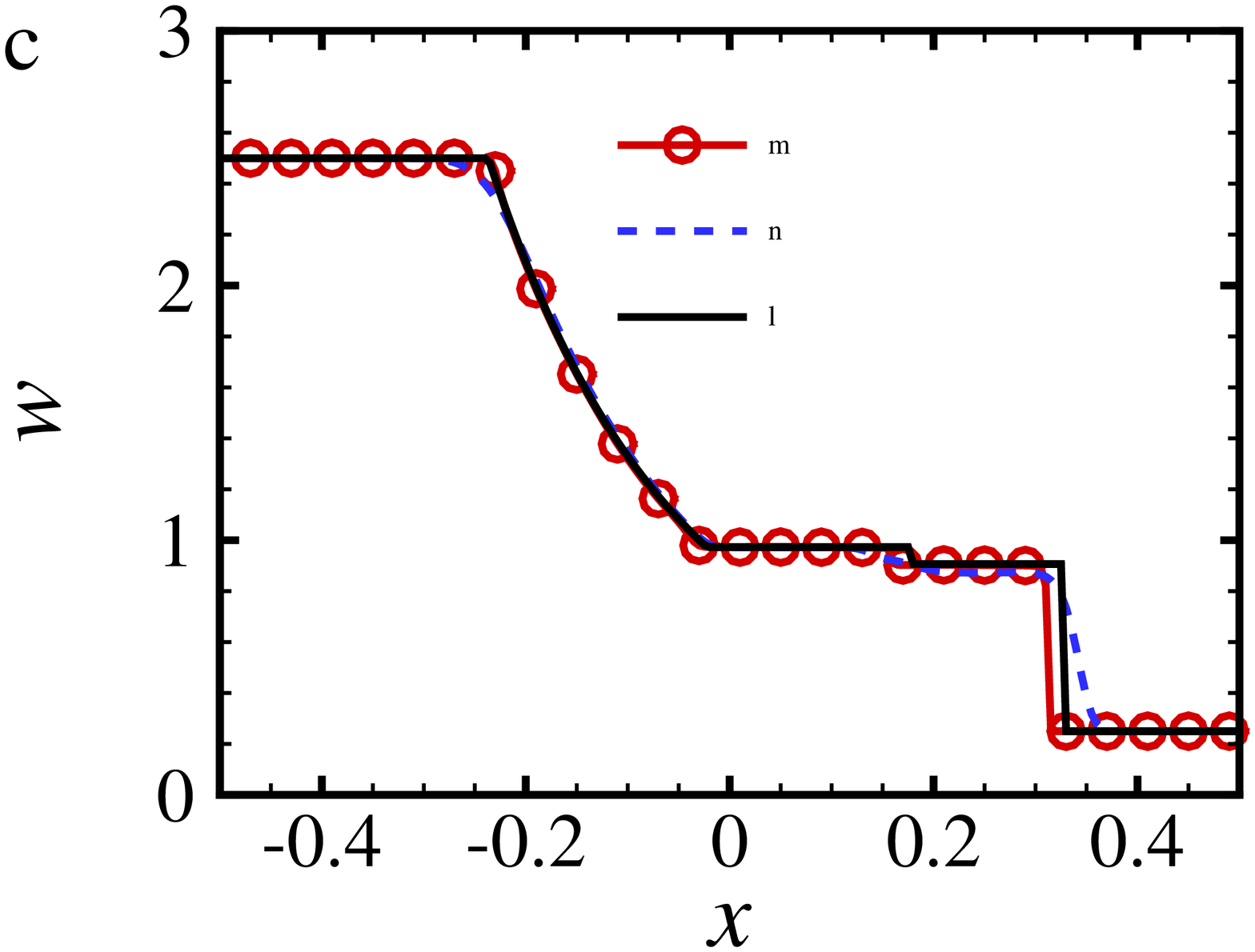}\\[0.5mm]
  \caption{Riemann problem with three components and nonlinear flux function. (a), (b), and (c) plot the exact solutions and the prediction results of the three components $u^{(1)}$, $u^{(2)}$, and $u^{(3)}$ using RoeNet and Roe solver.}
  \label{fig:nontrivial3c}
\end{figure}

We now access the performance of our model on solving Riemann problems with nonlinear flux functions, which is in the form of \eqref{eq:conserv_u} with $\bm F(\bm u) = \bm A(\bm u) \bm u$. Specifically, we apply our model to the Sod shock tube problem \cite{sod1978survey}, which is a one-dimensional Riemann problem in the following form
\begin{equation}
\begin{dcases}
\bm u = (\rho,\rho v, E)^T\\
\bm  F =  [\rho v , \rho v^2+p,v(E+p)]^T,\\
(\rho,p,v)|_{t=0,x\leq 0} =(1,1,0),~~(\rho,p,v)|_{t=0,x>0} = (0.125,0.1,0),
\end{dcases}
\label{eq:case3}
\end{equation}
where $\rho$ is the density, $p$ is the pressure, $E$ is the energy, and $v$ is the velocity. The pressure, $p$, is related to the conserved quantities through the equation of state
\begin{equation}
    p = (\gamma -1)\left(1-\frac12 \rho v^2\right)
\end{equation}
with $\gamma = 1.4$. The time evolution of this problem can be described by solving the Euler equations, which leads to three characteristics, describing the propagation speed of the various regions of the system. Namely the rarefaction wave, the contact discontinuity and the shock discontinuity \cite{sod1978survey}. 
In Figure \ref{fig:nontrivial3c}, we plot the three components of the problem. Similar to the conclusion drawn from the previous section \ref{sec:firstlinear}, RoeNet exhibits higher accuracy at predicting the discontinuities of the nonlinear Riemann problem.

\subsection{Predict discontinuity with smooth training data for inviscid Burgers' equation}
In this section, we exhibit the unique ability of our model to accomplish tacks that traditional machine learning approaches fail to complete.
Given a short window of continuous training data, we aim to use our model to predict long-term discontinuity of a nonlinear hyperbolic PDE, the inviscid Burgers' equation.
Burgers' equation is a fundamental PDE occurring in various areas, such as fluid mechanics, nonlinear acoustics, gas dynamics, and traffic flow. The inviscid Burgers' equation is a conservation equation, more generally a first order quasilinear hyperbolic equation, which can develop discontinuities (shock waves) \cite{burgers1948mathematical}. The set of equations is given by
\begin{equation}
\begin{dcases}
 F =  \frac12 u^2,\\
u(t=0,x) = \frac12+\sin(2\pi x).
\end{dcases}
\label{eq:case4}
\end{equation}

\begin{figure}
  \psfrag{x}[c][c]{\small $x$}
  \psfrag{u}[c][c]{\small $u$}
  \psfrag{m}{\small RoeNet}
  \psfrag{n}{\small Roe solver}
  \psfrag{a}{\small (a)}
  \psfrag{b}{\small (b)}
  \psfrag{c}{\small (c)}
  \centering
  \includegraphics[width=.333\linewidth]{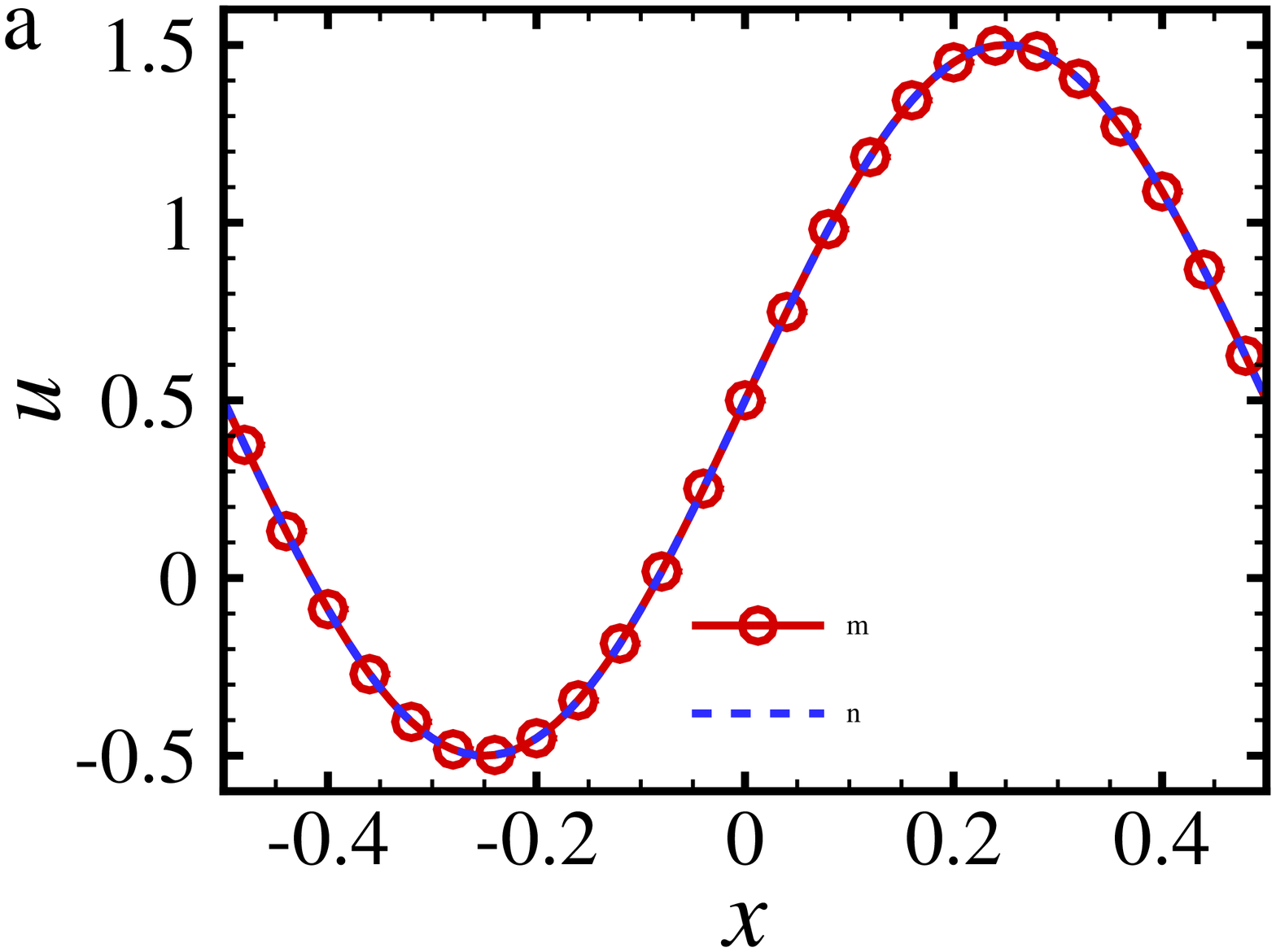}\hfill
  \includegraphics[width=.333\linewidth]{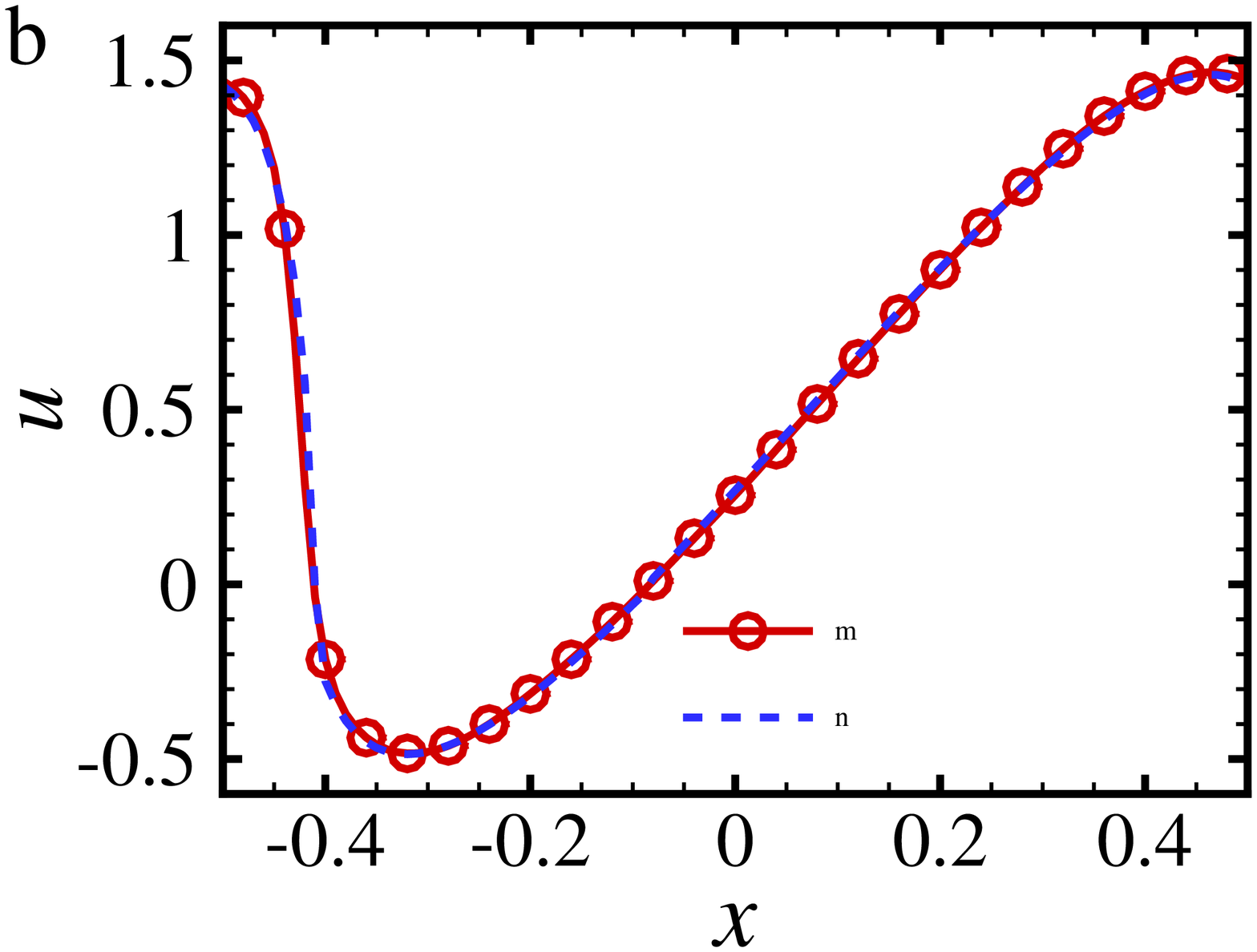}\hfill
  \includegraphics[width=.333\linewidth]{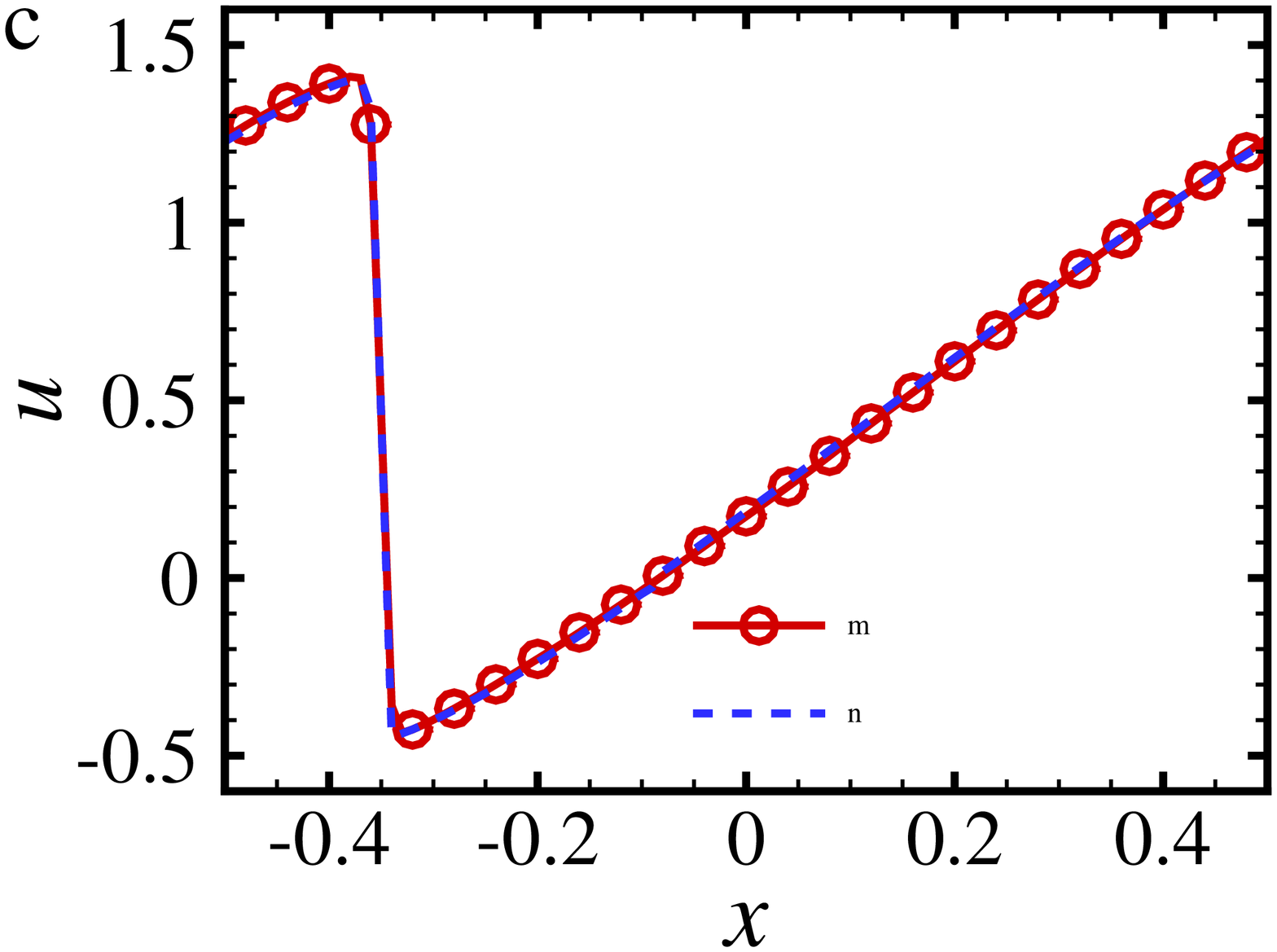}\\[0.5mm]
  \caption{Inviscid Burgers' equation with $u(t=0,x) = 0.5+\sin(2\pi x)$ at (a) $t = 0$, (b) $t =0.15$, and (c) $t=0.3$. Since there is no analytical solution for this problem, we plot only the prediction results made by RoeNet and Roe solver.}
  \label{fig:sint}
\end{figure}

Since there is no analytical solution for this problem, we plot only the prediction results made by RoeNet and Roe solver at $t= 0$, $t= 0.15$, and $t= 0.3$ in Figure \ref{fig:sint}. The perfect match of the predictions made by RoeNet with these made by Roe solver at all three time points shows that RoeNet successfully learn the future discontinuities of the problem based only on short-term continuous training data. This is a breakthrough improvement in solving prediction problems, as predicting long-term discontinuities from a short window of smooth training data is in general considered impossible using traditional machine learning approaches.

\section{Conclusion}

We presented here Roe Neural Networks (RoeNets) in solving HCLs \eqref{eq:conserv_u}. Our experiments show higher accuracy in stably predicting first-order linear hyperbolic PDEs with single or multiple components; RoeNet presents strong robustness against the introduction of arbitrary noise. In both experiments, RoeNet outperforms the traditional Roe solver, widely recognized as one of the most important modern high resolution, shock-capturing approximate Riemann solvers. The capacity to accurately predict discontinuities without dissipation is further demonstrated in the nonlinear Riemann setting. Uninformed of physical insights, RoeNet better captures the three physical characteristics: the rarefaction wave, the contact discontinuity, and the shock discontinuity. Remarkably, RoeNet approaches the Burgers' equation, a HCL that exhibits canonical discontinuities, in sufficient accuracy with limited training data that are strictly continuous. The accuracy and robustness attained are entirely attributed to our proposed network structure, thus makes contribution distinguishable from deep learning based solvers in which \emph{a priori} knowledge of solutions and data are encoded.

\section{Broader Impact}
This research marks a breakthrough improvement in solving HCLs by untilizing deep learning as a tool to solve the Roe matrix, which cannot be directly constructed by any general numerical method. The ability of our model to generate highly accurate predictions of evolution of convection without dissipation makes our model a better candidate in solving hyperbolic systems than the traditional numerical solvers. Moreover, our model makes significant advancement in solving the prediction problem, as it is capable of predicting long-term discontinuities from a short window of continuous training data, which is in general considered impossible using traditional machine learning approaches. Since HCLs can describe the behaviors of shock waves and rarefaction waves which are common in physical environments, our model enjoys broad applications in hydrodynamics, magnetohydrodynamics, aerodynamic, geophysics, and nuclear physics. This research does not bring any direct ethical consequence, but the application of our model to fields like aerodynamic, nuclear physics can potentially cause ethical issues.

\bibliographystyle{plain}
\bibliography{refs}

\end{document}